\documentclass[doublecol]{epl2}
\usepackage{amsmath}
\usepackage{amssymb}
\usepackage{mathtools}
\usepackage{txfonts}
\UseRawInputEncoding

\title{Pull-in features of nanoswitches in the Casimir regime with {\protect \\} account
of contact repulsion}
\shorttitle{Pull-in features of nanoswitches in the Casimir regime with account
of contact repulsion}

\author{G.~L.~Klimchitskaya\inst{1,2} \and A.~S.~Korotkov\inst{2}
\and V.~V.~Loboda\inst{2} \and V.~M.~Mostepanenko\inst{1,2,3}
 }
\shortauthor{G.~L.~Klimchitskaya \etal}

\institute{
  \inst{1}Central Astronomical Observatory at Pulkovo of the
Russian Academy of Sciences, Saint Petersburg,
196140, Russia\\
\inst{2}Peter the Great Saint Petersburg
Polytechnic University, Saint Petersburg, 195251, Russia\\
\inst{3}Kazan Federal University, Kazan, 420008, Russia
}

\abstract{The cantilever tip of a nanoswitch in close proximity to the ground plate
is considered with account of electrostatic, elastic, van der Waals (Casimir), and also
contact repulsive forces. The van der Waals (Casimir) and contact repulsive forces
are computed for a Si cantilever and either Au or Ni ground plates using the
Lifshitz theory and the method of pairwise summation with account of surface
roughness. It is shown that at short separations an impact of the van der Waals
(Casimir)
force leads to the pull-in and collapse of a cantilever onto the ground plate if the
contact repulsion is disregarded. Taking into consideration contact repulsion,
the nanoswitch is demonstrated to have the stable cyclic behavior with no pull-in
when switching voltage on and off.
}

\begin{document} \maketitle

\newcommand{\kb}{{k_{\bot}}}
\newcommand{\xl}{{i\xi_l}}
\newcommand{\ve}{{\varepsilon}}
\newcommand{\okt}{{(\omega,k_t)}}

\section{Introduction}
It is common knowledge that switches are employed to connect or disconnect electrical
circuits. For this reason, they are ``a must" instrument in both the scientific laboratory
and everyday life. During the last decades, the miniature switches called microswitches
find expanding applications because they are actuated by a relatively little force and have
rather small sizes. They are used in electrical appliances, machinery, communication,
navigation and radar systems, switching between reception and transmission modes,
and in many other cases. Microswitches are usually operated by the mechanical restoring
and electrostatic attractive forces (see, for instance, the review papers \cite{1,1a,2,3,4,5,6,7}).
With increasing voltage, the magnitudes of both forces increase. At some value of voltage,
however, the pull-in instability of a switch happens when the mechanical elastic force
cannot counterbalance the electrostatic force any more leading to device collapse \cite{1a,2,7}.
Knowledge of the conditions under which this phenomenon occurs is important for both the
production and operation of microswitches \cite{8,9}.

With further decrease of characteristic sizes of microswitches, the separation distances
between static electrodes and moving beams fall to below a micrometer. In this case,
apart from the classical elastic and electrostatic forces, the additional forces of quantum
nature come into play. The most well known of them are the van der Waals and Casimir
forces caused by the zero-point and thermal fluctuations of the electromagnetic field
\cite{10,11,12,13,14}. These are in fact the manifestations of the same force
at separations below 2-3 nm and at larger separations, where the relativistic effects become
important, respectively. In the distance range of several hundred nanometers, the Casimir force is
comparable in size with the characteristic electric force and at separations below 100 nm far
exceeds it. There is an extensive literature on the role of van der Waals and Casimir forces in functioning
of various nanodevices \cite{15,16,17,18,19,20,21,22,23} and their pull-in instability \cite{24,25,26}.

Microswitches with reduced separations between the cantilever beam and ground plate are
usually called nanoswitches. For nanoswitches, the pull-in voltage, when the beam becomes
unstable and collapses onto the ground plate, is determined by the combined action of
the electrostatic, $F_{\rm el}$, elastic, $F_{\rm elas}$, and van der Waals and Casimir forces,
$F_{\rm vdW}$. The prerequisites to the formation of the pull-in in nanoswitches were investigated
in several papers. Thus, in \cite{2,27,28,29,30,32,33,34,35} the bending of the nanoswitch
beam under an impact of the elastic, electrostatic and Casimir forces was described in the
framework of Euler-Bernoulli beam theory, whereas for the Casimir force the
simplified expression valid for two parallel plates made of ideal metal has been used. In
\cite{36,37,38,39} the role of the Casimir force in nanoswitches was investigated
with account of surface roughness. For this purpose, the material of the plates was also
considered as perfectly conducting and deflection of the cantilever beam was described by the
Hooke's law. The cases of both zero and nonzero temperature were considered.

It should be taken into account that, under an impact of the Casimir force, the
pull-in phenomenon in nanoswitches occurs at shorther separations and becomes more
complicated. The point is that it is not easy to control the value of this force. If, for example,
the pull-in leading to a collapse of a nanoswitch happens at some critical voltage, one can
simply apply only lesser voltages. The value of the Casimir force, however, is fully
determined by the plate materials, temperature, and the separation distance. Thus, to decrease
the value of this force in the nanoswitch by 20 to 25\%, it was suggested \cite{40} to use
configuration of an Au sphere and an alloy of silver, indium, antimony, and tellurium deposited
on an Al-coated silicon plate. When heated by a laser, this alloy reversibly transforms from a
crystalline to an amorphous state decreasing the Casimir force value. The optical switch
exploiting the variation of the Casimir force in sphere-plate geometry under an impact of light
was also suggested \cite{41}.

All the above-listed articles considering the role of the Casimir force in nanoswitches
described it in the first approximation by the original Casimir expression valid for ideal metal
surfaces. This idealization, however, does not work at short separation distances characteristic
for nanoswitches. In two more articles, which are devoted to the role of Casimir
force in the optical switch using a graphene sheet \cite{42} and to the silicon carbide nanoswitch
\cite{43}, this force was calculated in the framework of the Lifshitz theory \cite{11,12,13,14} taking
into account the dielectric permittivities of all invloved materials over the wide frequency
regions. Thus, in \cite{43} the force between artificially corrugated contact surfaces,
designed to reduce the contact area, was computed at separations down to 0.1 nm using the
Lifshitz formula and the Derjagiun approximation \cite{14}.

A point that should be mentioned at this place is the application area of the Lifshitz theory.
As a semiclassical theory, it considers the quantized electromagnetic field but continuum classical
plate materials. Such a theory is applicable at separations between the plates which are much
larger than interatomic separations \cite{11,12,13}, i.e., much larger than 0.3-0.4 nm. What
this means is Lifshitz theory cannot be reliably applied at separations below 3 nm. What's more,
at separations below a few angstroms between the surfaces the quantum repulsive forces of
exchange character take effect. These forces act between the closest atoms and molecules of
rough contact surfaces \cite{44} and prevent the cantilever beam from collapsing onto the
ground plate, but they were not yet addressed in the literature devoted to the pull-in
features of nanoswitches working in the Casimir regime.

In this letter, the nanoswitch, which can reset a device from, e.g., reception to transmission mode
by applying the voltages of opposite sign between a moving  Si cantilever beam and two fixed
metallic plates, is considered with account of the electrostatic, elastic, Casimir, and
contact repulsive forces. Keeping in mind that the overall process of cantilever bending under an
impact of elastic and electrostatic forces was already considered in the literature in sufficient detail
\cite{2,27,28,29,30,32,33,34,35}, we concentrate on the analysis of its behavior in close
proximity to one of the plates where the elastic force is almost constant, but the forces of
quantum nature should be accounted for more accurately.

The Casimir and contact repulsive forces are calculated for the cantilever made
of conductive (doped) Si and either Au or Ni plates. At separations above 3 nm the Casimir force is
computed by means of the Lifshitz theory. At shorter separations, the van der Waals force is calculated
by the additive summation of interatomic van der Waals potentials with subsequent normalization
of the obtained results by the factor, for which the additive summation exceeds the Lifshitz
theory at 3 nm. The contact repulsion is found by the additive summation of the interatomic
repulsive potentials. It is suggested that the surface roughness is the same as in experiments
on measuring the Casimir force between Au \cite{45}, Ni \cite{46,47,48} and Si \cite{49,50}
surfaces by means of an atomic force microscope.

According to our results, with account of contact repulsion, the nanoswitch remains efficient
within a rather wide range of applied voltages. Thus, owing to contact repulsion, it becomes
possible to expand the application conditions of this nanoswitch avoiding the negative
consequences of its pull-in instability.

\section{Casimir force at separations above 3\,nm}
We consider the cantilever beam of 120 $\mu$m height and 10 $\mu$m diameter arranged at
mid-position between two plates made of either Au or Ni spaced at about 30 $\mu$m apart
(see fig. 1(a) not to scale and the micrograph of a cantilever in fig. 1(b)). The potential difference $U$
applied between the cantilever and the left plate gives rise to electrostatic force $F_{\rm el}$, which
causes cantilever bending. The process of bending under an impact of electrostatic 
force and the opposite
in sign elastic force is consdered in \cite{2,27,28,29,30,32,33,34,35}. Here, we concentrate our
attention on the position of a cantilever tip in close proximity to the plate where the additional
forces of quantum origin come into play. These are the Casimir and exchange repulsive
forces. They are completely determined by the interaction between the spherical tip of a cantilever
and metallic plate (see fig. 2).

\begin{figure}[!h]
\onefigure{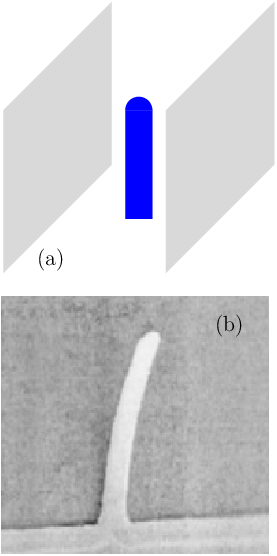}
\caption{The nanoswitch made up of a Si cantilever and two parallel Au or Ni plates. (a) Schematic picture and
(b) micrograph of the cantilever bending. }
\label{fig.1}
\end{figure}

\begin{figure}
\onefigure{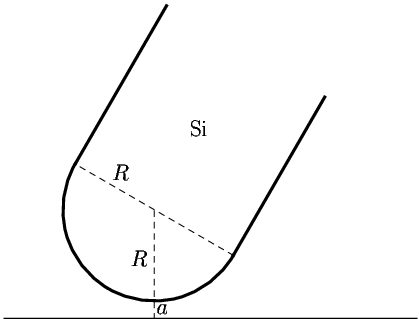}
\caption{Schematic picture of the spherical tip of a cantilever at the closest separation $a$ from the plate.}
\label{fig.2}
\end{figure}

If the closest separation $a$ between the smooth surfaces of the plate and the spherical cantilever tip
slightly exceeds 3 nm, the Casimir force acting between them can be calculated by the
Lifshitz formula used in the proximity force approximation \cite{14}
\begin{eqnarray}
&&
F_{\rm vdW}^L(a)=k_BTR\sum_{l=0}^{\infty}{\vphantom{\sum}}^{\prime}\!\!
\int_{0}^{\infty}\!\!\!k_{\bot}dk_{\bot}
\label{eq1} \\
&&~~
\times
\left\{\ln[1-r_{{\rm TM},l}^{(1)}r_{{\rm TM},l}^{(2)}e^{-2aq_l}]
+\ln[1-r_{{\rm TE},l}^{(1)}r_{{\rm TE},l}^{(2)}e^{-2aq_l}] \right\},
\nonumber
\end{eqnarray}
\noindent
where $k_B$ is the Boltzmann constant, $T$ is the temperature, $R$ is the sphere
radius,  the prime on the summation sign divides the term with $l=0$ by 2,
$k_{\bot}$ is the magnitude of the wave vector
projection on the plane of a plate, $\xi_l=2\pi k_BTl/\hbar$ with $l=0,\,1,\,2,\,\ldots$
are the Matsubara frequencies, and
$q_l^2=k_{\bot}^2+\xi_l^2/c^2$. The force (\ref{eq1}) is negative, i.e., attractive.

The reflection coefficients for two polarizations of the electromagnetic field,
transverse magnetic (TM) and transverse electric (TE), have the form
\begin{eqnarray}
&&
r_{{\rm TM},l}^{(p)}=r_{\rm TM}^{(p)}(i\xi_l, k_{\bot})=
\frac{\varepsilon_l^{(p)}q_l-k_l^{(p)}}{\varepsilon_l^{(p)}q_l+k_l^{(p)}},
\nonumber \\
&&
r_{{\rm TE},l}^{(p)}=r_{{\rm TE},l}(i\xi_l,k_{\bot})=
\frac{q_l-k_l^{(p)}}{q_l+k_l^{(p)}},
\label{eq2}
\end{eqnarray}
\noindent
where ${k_l^{(p)}}^2=k_{\bot}^2+\varepsilon_l^{(p)}\xi_l^2/c^2$ and
$\varepsilon_l^{(p)}=\varepsilon^{(p)}(i\xi_l)$ are the dielectric permittivities of
a plate ($p$=1) and a sphere ($p$=2) materials. In effect, only the closest to a plate
part of the sphere determines the total cantilever-plate interaction in the position
of fig. 2.

Note that the proximity force approximation states that $F_{\rm vdW}^L=2{\pi}RE_{\rm vdW}^L$,
where $E_{\rm vdW}^L$ is the free energy per unit area of two plane parallel plates. It is well
applicable in our case because $a$ is below a few nanometers and $R=5 \mu$m, so that
$a/R \ll 1$ \cite{51}. The dielectric permittivities of Au, Ni, and Si along the imaginary
frequency axis are computed by means of the Kramers-Kronig relation using the tabulated
optical data for these materials \cite{52}. The computational results obtained by eqs.
(\ref{eq1}) and (\ref{eq2}) in the region of separations from 3 to 5 nm at $T=300$K are
presented in columns 2 and 3 of table 1 for a Si cantilever and either Au or Ni plate,
respectively.
\begin{table}[h]
\caption{The values of the Casimir force  between a Si cantilever and either
an Au or Ni plates are shown in columns 2 and 3, respectively, as a function of separation.}
\label{tab1}
\begin{center}
\begin{tabular}{cll}
$~~a$~(nm)\hspace*{2mm}&\multicolumn{2}{c}{$|F_{\rm vdW}^L|$~(nN)}\\
& Si-Au \hspace*{5mm}& Si-Ni \\
~3.0& 28.32& 26.86\\
~3.5& 20.03 & 19.00 \\
~4.0& 14.83 & 14.07\\
~4.5& 11.36 & 10.78\\
~5.0& 8.95 & 8.50\\
\end{tabular}
\end{center}
\end{table}

In reality, the surfaces of interacting bodies are not smooth but covered by roughness
which contributes essentially to the force at short separations. The effect of
roughness can be taken into account by the method of geometrical averaging \cite{14,53}.
For this purpose, the roughness profiles on both surfaces are measured by means of an
atomic force microscope. The obtained data are used to find the relative fractions $p_i$
of the contact surfaces which are separated by the distances $a_i$ ($i=1,2,...,N$). As a result,
the force between rough surfaces is calculated as
\begin{equation}
F^{(r)}(z)=\sum_{i=1}^{N}p_iF(a_i),
\label{eq3}
\end{equation}
\noindent
where $z$ is the separation distance between the mean levels of the roughness profiles on
both surfaces. If $F(a_i)=F_{\rm vdW}^L(a_i)$ calculated by eqs. (\ref{eq1}) and (\ref{eq2})
valid for the prefectly smooth surfaces, eq.  (\ref{eq3}) allows calculation of the van der Waals
(Casimir) force $F_{\rm vdW}^{(r)}$ between the rough surfaces. However, eq.  (\ref{eq3}) is
applicable also to the forces of different nature (see below).

It is significant that even if $z \ge 3$ nm there are $a_i$ which are smaller and larger than 3 nm.
In doing so, for $a_i < 3$ nm the van der Waals (Casimir) force $F_{\rm vdW}(a_i)$, strictly speaking,
cannot be calculated by eqs. (\ref{eq1}) and (\ref{eq2})  because these $a_i$ are outside the
application region of the Lifshitz theory. The quantum forces at the shortest separations between the
surfaces are considered in the next section.

\section{van der Waals and exchange forces at the shortest separations}
At separations below 3 nm the interacting surfaces cannot be already considered as continuous
media, but rather as consisting of separate atoms or molecules. In this separation region, the
relativistic effects do not play any role and the Casimir force coincides with
the standard van der Waals interaction. In the case of two similar atoms, it is described by the
nonretarded potential \cite{54,55,56,57,58,59,61}
\begin{eqnarray}
&&
V(r)=-4\epsilon\left(\frac{\sigma}{r}\right)^6,
\label{eq4}
\end{eqnarray}
\noindent
where $r$ is the interatomic separation and $\epsilon$ and $\sigma$ are the constants specific for
each atom (molecule). The values of $\epsilon$ and $\sigma$ are determined by the methods of
molecular dynamics. For two different atoms characterized by the parameters $\epsilon_1$, $\sigma_1$
and $\epsilon_2$, $\sigma_2$ the parameters $\epsilon_{12}$, $\sigma_{12}$ of their interaction
are obtained by the Lorentz-Berthelot combination rules \cite{54,57,59}
\begin{eqnarray}
&&
\epsilon_{12}=\sqrt{\epsilon_1\epsilon_2},  \quad \sigma_{12}=\frac{1}{2} (\sigma_{1}+\sigma_{2}).
\label{eq5}
\end{eqnarray}

The values of the interaction parameters for atoms of our interest and their pairs are
presented in table 2 \cite{58,59,61}.
\begin{table}[h]
\caption{Parameters of the van der Waals interaction potential  between the atoms of Au, Ni,
Si and their pairs are presented in columns 2, 3, 4, 5, and 6, respectively.}
\label{tab2}
\begin{center}
\begin{tabular}{lccccc}
\hspace*{3mm}& Au \hspace*{3mm}& Ni \hspace*{3mm}& Si \hspace*{3mm}& Si-Au \hspace*{3mm}& Si-Ni\\
~$\epsilon\times10^{20}\,$(J)& 7.15& 8.355& 0.279& 1.412& 1.526\\
~$\sigma\times10^{10}\,$(m)& 2.638& 2.275& 3.826& 3.232& 3.050\\
\end{tabular}
\end{center}
\end{table}

Now we calculate the interaction energy due to potential  (\ref{eq4}) between an atom of Si and a semispace
made of either atoms of Au or Ni (this energy does not depend on the plate thickness if it exceeds 100 nm
\cite{14}). The atom of Si is at the height $z$ above a semispace. Integrating the potential  (\ref{eq4})
over the semispace volume (the index 1 is for Si and the index 2 for Au or Ni), one obtains
\begin{eqnarray}
&&
V_{12}(z)=-8\pi\epsilon_{12}\sigma_{12}^6n_2\int_{0}^{\infty}\!\!\!\rho d\rho
\int_{-\infty}^{0}\!\frac{dz_2}{[(z-z_2)^2+\rho^2]^3}
\nonumber \\
&&~~~~~~~~~
=-\frac{2}{3}\pi\epsilon_{12}\sigma_{12}^6n_2
\frac{1}{z^3},
\label{eq6}
\end{eqnarray}
\noindent
where $n_2$ is the density of Au or Si atoms per unit volume ($n_{\rm Au}\approx 5.907\times10^{28}$
m$^{-3}$, $~~n_{\rm Ni}\approx 9.138\times10^{28}$m$^{-3}$).

Integrating eq. (\ref{eq6}) over the Si semispace spaced at the separation $a$ from the metallic one,
we find their interaction energy per unit area
\begin{equation}
\frac{E_{12}}{S}=-\frac{2}{3}\pi\epsilon_{12}\sigma_{12}^6n_1n_2\int_{a}^{\infty}\!\!\frac{dz}{z^3}
=-\frac{1}{3}\pi\epsilon_{12}n_1n_2\sigma_{12}^4\left(\frac{\sigma_{12}}{a}\right)^2,
\label{eq7}
\end{equation}
\noindent
where $n{_1}\approx 4.996\times10^{28}$m$^{-3}$ is the number of Si atoms per unit volume.

Using the proximity force approximation, we now obtain the additive expression for the van der Waals
force between a Si sphere and either an Au or a Ni plate
\begin{equation}
F_{\rm vdW}^{\rm add}(a)=2\pi R\frac{E_{12}}{S}=-\frac{2}{3}\pi^2R\epsilon_{12}n_1n_2\sigma_{12}^4
\left(\frac{\sigma_{12}}{a}\right)^2.
\label{eq8}
\end{equation}

It has been known \cite{14,62,63} that the additive summation of interatomic potentials overestimates the
value of the van der Waals force between macroscopic bodies. To partially take into account the effects
of nonadditivity, it was suggested
to divide the result  (\ref{eq8}) by the normalization factor equal to the ratio of the additive expression for
the force to the exact Lifshitz expression at the separation distance where the latter is yet applicable
(i.e., at $a=3$~nm) \cite{14,62,63}
\begin{equation}
K=\frac{F_{\rm vdW}^{\rm add}(a=3\rm ~nm)}{F_{\rm vdW}^{L}(a=3\rm~ nm)},
\label{eq9}
\end{equation}
\noindent
where $F_{\rm vdW}^{L}$ is defined by eq. (\ref{eq1}).

Then the corrected van der Waals force at separations below 3~nm is given by
\begin{equation}
F_{\rm vdW}^{\rm corr}(a)=-\pi^2 R\epsilon_{12}n_1n_2\frac{2\sigma_{12}^4}{3K}\left(\frac{\sigma_{12}}{a}\right)^2.
\label{eq10}
\end{equation}
\noindent
This expression is the exact one at $a=3$~nm. It provides a good approximation for the van der Waals force
at shorter separations.

In fig. 3, the magnitudes of the van der Waals force between the Si top of a cantilever and Au and Ni plates
are shown as functions of separation over the region from 0.1 to 5~nm. For $3~\mbox{nm} \leq a \leq5$~nm, the
force values were computed by eq. (\ref{eq1}), see table 1. For $0.1~\mbox{nm} \leq a \leq3$~nm, eq. (\ref{eq10})
was used in computations. In the main figure field, a difference between the lines plotted for the Au and Ni
plates is indistinguishable. It is clearly seen, however, in the inset plotted on an enlarged scale.
\begin{figure}[!b]
\onefigure{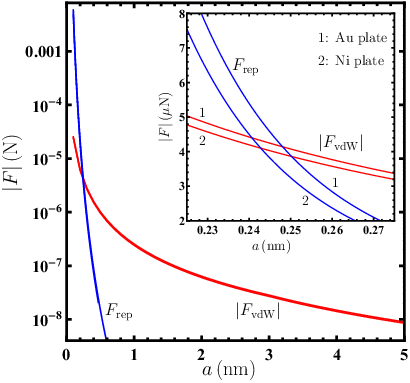}
\caption{The magnitudes of the van der Waals (Casimir) and contact repulsive forces between the smooth
surfaces of a Si cantilever tip and either Au or Ni plates are shown as functions of separation.
The region of short separations is shown in the inset on an enlarged scale. }
\label{fig.3}
\end{figure}

Between the atoms of a Si cantilever tip and metallic plates separated by a few angstroms, there are also
the quantum repulsive forces of exchange nature. Their interaction potential is usually parametrized
as \cite{54,55,56,57,58,59,61}
\begin{equation}
V_{\rm rep}(r)=4\epsilon\left(\frac{\sigma}{r}\right)^{12}.
\label{eq11}
\end{equation}
\noindent
Note that the sum of potentials (\ref{eq4}) and (\ref{eq11}) is the familiar Lennard-Jones
potential \cite{54,55,56,57,58,59,61}.

By repeating the same calculation as in eqs. (\ref{eq6})--(\ref{eq8}), but with the potential (\ref{eq11})
in place of (\ref{eq4}), one obtains an expression for the contact repulsive force between a Si
cantilever and Au or Ni plates
\begin{equation}
F_{\rm rep}(a)=\frac{1}{45}\pi^2R\epsilon_{12}n_1n_2\sigma_{12}^4\left(\frac{\sigma_{12}}{a}\right)^8.
\label{eq12}
\end{equation}

In fig. 3, the contact repulsive force (\ref{eq12}) is shown as the function of separation. Similar to the
van der Waals force, the values of repulsive force between a Si cantilever tip and either an Au or 
a Ni plate
are almost coinciding. On an enlarged scale of the inset, the difference between the cases of Au and
Ni plates is clearly seen.

In the end of this section, we also present an expression for the classical electrostatic force between the
cantilever tip and metallic plate. In the framework of the proximity force approximation, it is given by
\begin{equation}
F_{\rm el}(a)=-\frac{\pi R\varepsilon_0U^2}{a},
\label{eq13}
\end{equation}
\noindent
where $\varepsilon_0$ is the permittivity of free space.

\section{Forces between rough surfaces and pull-in instability}
As was noted above, at short separations the surface roughness may essentially contribute to the
force value. In fact, the roughness profiles should be examined by means of an atomic force
microscope for each specific sample. Here, we assume that the roughness on Au and Ni plates
is chatacterized by the same r.m.s. variances $\delta_{\rm Au}=2$~nm and $\delta_{\rm Ni}=1.5$~nm
as in the experiments on measuring the Casimir force \cite{45} and \cite{46,47,48}, respectively.
The Si surface is assumed to be perfectly smooth (according to the experiment \cite{49,50},
$\delta_{\rm Si}=0.1$~nm, i.e., is very small). The absolute separation $z$ between the surfaces
is reckoned from the mean level of a roughness profile on the plate. For the sake of simplicity,
below we assume that with a probability of 50\% the points of a plate and of a cantilever are at
the separation distances of $z+\delta_{\rm Au,Ni}$ and $z-\delta_{\rm Au,Ni}$. For this simple
model of roughness, the mean level is at the height $\delta_{\rm Au,Ni}$ above a plate.

In this case, eq. (\ref{eq3}) leads to the following expression for the van der Waals (Casimir)
force acting in our nanoswitch with account of surface roughness
\begin{equation}
F^{(r)}_{\rm vdW}(z)=\frac{1}{2}[F_{\rm vdW}^{L,\rm corr}(z+\delta_{\rm Au,Ni})+
F_{\rm vdW}^{L,\rm corr}(z-\delta_{\rm Au,Ni})].
\label{eq14}
\end{equation}
\noindent
When calculating $F^{(r)}_{\rm vdW}(z)$, it is necessary to use eq. (\ref{eq1}) for
the functions $F_{\rm vdW}^{L,\rm corr}$ if their arguments exceed 3~nm and eq. (\ref{eq10})
if the arguments are below 3~nm.

Similar to eq. (\ref{eq14}) expressions follow from eq. (\ref{eq3}) for the contact repulsive and
electric forces with account of surface roughness
\begin{equation}
F^{(r)}_{\rm rep,el}(z)=\frac{1}{2}[F_{\rm rep,el}(z+\delta_{\rm Au,Ni})+
F_{\rm rep,el}(z-\delta_{\rm Au,Ni})],
\label{eq15}
\end{equation}
\noindent
where $F_{\rm rep}(a)$ and  $F_{\rm el}(a)$ are given by eqs. (\ref{eq12}) and (\ref{eq13}),
respectively.

Let us now discuss the pull-in features of the nanoswitch under consideration in the position of
its cantilever in close proximity to the contact. In \cite{27,29,43} the equilibrium state of a
cantilever before it turns back to the central position when the voltage is switched off was
found from the equation
\begin{equation}
F_{\rm el}(z)+F_{\rm vdW}(z)+F_{\rm elas}(z)=0,
\label{eq16}
\end{equation}
\noindent
where the electric and van der Waals (Casimir) forces are negative and the elastic force is positive.
Note, however, that this equilibrium state is unstable because any inavoidable small displacement
of the cantilever tip to shorter separation due to thermal noise would result in an increased van der Waals
attraction and pull-in adhering of the cantilever to the plate.

\begin{figure}[!b]
\onefigure{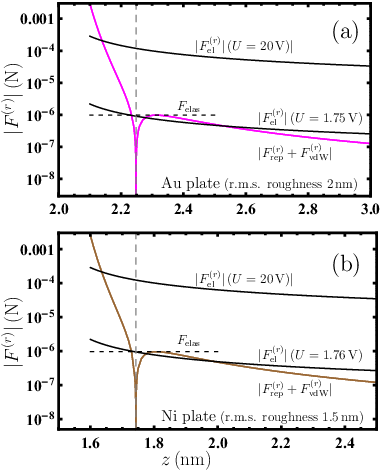}
\caption{The magnitudes of the elastic, electric (at the applied voltages of (a) 1.75, 
 (b) 1.76, and 20~V) and the sum of the
van der Waals (Casimir) and contact repulsive forces between the surface of a Si cantilever and
rough  surfaces of (a) Au and  (b) Ni plates are shown as functions of separation.}
\label{fig.4}
\end{figure}
With account of the repulsive force, the situation reverses. In fig. 4(a), we plot the magnitude of the sum
of repulsive and van der Waals forces calculated between the rough surfaces of a Si cantilever tip and an Au
plate by eqs. (\ref{eq14}) and (\ref{eq15}) as a function of separation. It is seen that this quantity
reaches the maximum value of 0.978 $\mu$N at $z=2.31$~nm. Taking into account that to the right of the
vertical dashed line in fig. 4(a) the force $F^{(r)}_{\rm rep}+F^{(r)}_{\rm vdW}$ is attractive, i.e.,
negative, it reaches the stable minimum at $z=2.31$~nm.

At the shortest separations below several nanometers, the restoring elastic force $F_{\rm elas}$ is
essentially constant \cite{43}. If its value satisfies the condition
\begin{equation}
F_{\rm elas} \geq {\rm max}~ |F_{\rm rep}(z)+F_{\rm vdW}(z)|=0.978~\mu \rm N,
\label{eq17}
\end{equation}
\noindent
then, by switching the electric force off, the cantilever is brought back to its original position between
the plates (see fig. 1(a)). The minimum restoring elastic force is shown in fig. 4(a) by the
horizontal dashed line. The value of this force is in agreement with that found in \cite{43}.

The question arises what is the value of voltage $U$ which brings the cantilever tip to close
proximity of the plate surface. This value can be estimated from the equilibrium position of a
cantilever at the separation $z$ where the force $F_{\rm elas}$ is still constant
\begin{equation}
F^{(r)}_{\rm el}(z)+F^{(r)}_{\rm vdW}(z)+F^{(r)}_{\rm rep}(z)+F_{\rm elas}(z)=0.
\label{eq18}
\end{equation}

Thus, substituting in eq. (\ref{eq18}) the values of $F^{(r)}_{\rm el}$ and $F^{(r)}_{\rm vdW}$ at
$z=2.5$~nm (see fig. 4(a)), one obtains $F^{(r)}_{\rm el}(z=2.5~{\rm nm})=-0.473~\mu$N
which results from $U=1.75$~V according to eqs. (\ref{eq13}) and (\ref{eq15}). The electrostatic
force determined by this voltage is shown in fig. 4(a) as a function of separation.

Note that, similar to the case of eq. (\ref{eq16}), the separation of $z=2.5$~nm is the position of
an unstable equilibrium. After reaching it, however, the cantilever does not crash under an impact
of the attractive van der Waals force, but reaches the equilibrium position at $z=2.31$~nm under the
influence of the sum of forces $F^{(r)}_{\rm vdW}+F^{(r)}_{\rm rep}$. After the electric force
is switched off, the cantilever returns to its initial position. When the voltage of the same magnitude
but of the opposite sign is applied between the cantilever and the right plate, the nanoswitch takes a
system to the second state. After switching off the voltage, the cantilever again returns to the
initial position, and the cycle is closed.

It is significant that, with account of contact repulsion, the stable work of the nanoswitch with no
pull-in takes place in the wide range of larger applied voltages. For instance, the magnitude of the
electrostatic force corresponding to the applied voltage of $U=20$~V is shown in fig. 4(a) as a
function of separation. In this case, the contact position of the closest approach holds at
$a=2.135$~nm separation between the Si cantilever tip and the mean level of roughness on the
Au plate.

Similar calculations have been performed for the case of a Ni plate. The obtained results are
presented in fig. 4(b). The main difference of the case of Ni plate is that the maximum value of
the quantity $|F^{(r)}_{\rm rep}+F^{(r)}_{\rm vdW}|$ equal to 0.967~$\mu$N is now reached at
$z=1.81$~nm. As a result, the restoring elastic force in the proximity of the contact position should be
no less than 0.967~$\mu$N. The electric force determined at $z=2.0$~nm corresponds to
$U=1.76$~V voltage. With much larger applied voltage of 20~V, the closest separation between the
cantilever tip and the mean level of the roughness on a Ni plate is equal to $z=1.65$~nm. All these
differences are determined by the smaller r.m.s. roughness on a Ni plate. Note that an employment
of Ni as a plate material opens the possibility of using the magnetic forces in additions to the
electric ones in operating a nanoswitch.

\section{Conclusions and discussion}
In this letter, we have considered the cantilever of a nanoswitch in close proximity to the ground plate
where it comes under the influence of the van der Waals (Casimir) and contact repulsive forces in addition
to the electrostatic and elastic interactions. The cases of a cantilever made of conductive (doped) Si
interacting with either an Au or Ni plates were explored. The van der Waals, contact repulsive, and
electrostatic forces acting on a cantilever tip were computed with account of surface roughness.
In doing so, at separation distances exceeding 3~nm the Casimir force was computed
by means of the Lifshitz theory but at shorter separations the van der Waals force was found using
the method of the additive summation of nonretarded interatomic potentials corrected for nonadditivity.

According to our results, the equilbrium position of a nanoswitch working in the Casimir regime
in the absence of contact repulsion would be unstable leading to the pull-in collapse of the cantilever on
the ground plate. However, the account of repulsive force of exchange character gives rise to the stable
minimum in the sum of the van der Waals and contact repulsive forces resulting in the cyclic
functioning of a nanoswitch between two extreme positions with no pull-in phenomena. The nanoswitch
of this kind is operable within rather wide ranges of the electrostatic and elastic forces.

The obtained results may be useful when elaborating nanoswitches with further reduced dimensions and
contact areas for application in nanoelectronic devices of next generations.

\acknowledgments
The work of G.L.K. and V.M.M. was partially funded by the
Ministry of Science and Higher Education of Russian Federation
as part of the World-Class Research Center program: Advanced Digital Technologies
(contract No. 075-15-2022-311 dated April 20, 2022).
A.S.K. and V.V.L. thank the State Assignment for Basic Research
(project FSEG-2023-0016) for financial support.
The work of
V.M.M.~was also partially carried out in accordance with the Strategic
Academic Leadership Program "Priority 2030" of the Kazan Federal
University.
\vspace*{2mm}

{\it Data availability statement:}
All data that support the findings of this study are included within the article (and
any supplementary files).

\end{document}